\documentstyle[12pt,fleqn,espcrc1,epsfig]{article}

\newcommand{\AmS}{{\protect\the\textfont2
  A\kern-.1667em\lower.5ex\hbox{M}\kern-.125emS}}

\newcommand{\vk}{{\vec k}}

\newcommand{\vb}{{\vec b}}
\newcommand{\vp}{{\vec p}}
\newcommand{\vq}{{\vec q}}

\newcommand{\vC}{{\vec C}}
\newcommand{\vH}{{\vec H}}
\newcommand{\vB}{{\vec B}}
\newcommand{\tr}{{{\rm Tr}}}

\title{Jet quenching in thin plasmas} 

\author{M. Gyulassy$^a$, 
{\underline {\rm P. L\'evai}}$^{a,b}$, I. Vitev\address{
${}$ Department of Physics, Columbia University, New York, NY, 10027, USA \\
${}^b$ Research Institute for Particle and Nuclear Physics, POB 49,
       Budapest, 1525, Hungary} } 
       
\begin{document}

\maketitle

\begin{abstract}
We investigate the  energy loss of quarks and gluons produced in hard
processes resulting from final state rescatterings in a finite quark-gluon
plasma.  The angular distribution of the soft gluon brems\-strah\-lung
induced by $n_s=1$ rescatterings in the plasma is computed in
the Gyulassy-Wang model. Special focus is on how the
interference between the initial hard radiation amplitude, the multiple
induced Gunion-Bertsch radiation amplitudes, and gluon rescattering
amplitudes modifies the classical parton cascade results.
\end{abstract}

\section{INTRODUCTION}
At RHIC energies $(\surd s \sim 200$ AGeV) a new observable was 
predicted in nuclear collisions: jet quenching~\cite{BJ}-\cite{MGXW92}.
Collisional
energy loss of a hard jet propagating through quark-gluon plasma has been
estimated to be modest~\cite{BJ,THOMA}\/: $dE_{coll}/dx \ll 1$ GeV/fm, 
however the induced radiative energy loss
estimated~\cite{MGMP,GPTW} from the Gunion-Bertsch amplitude~\cite{GUNION}
is expected to be significantly larger, $dE_{rad}/dx> {\rm few}\; GeV/fm$. 
In Ref.~\cite{MGXW} a model for non-abelian energy
loss was developed and the analog of the QED class of radiation diagrams was
summed. This leads to a constant $dE/dx \sim 1$ GeV/fm independent of the path
length.  However, in BDMPS~\cite{BDMPS} it was shown that gluon
rescattering diagrams in the medium considerably increased the 
energy loss.  For an incident very high energy
jet penetrating a plasma of thickness, $L$, in which the mean free path,
$\lambda=1/(\sigma\rho)$, and the color electric fields are screened on a scale
$\mu$, the analytic BDMPS prediction \cite{BDMPS} is
${dE}/{dx}\propto \alpha_s \mu^2 {L}/{\lambda}$.
An alternative path integral approach  led to similar results~\cite{Zakharov}.
The transverse momentum dependence of the energy loss in QED was calculated
recently \cite{UrsMik99}.

Unfortunately, the analytic approximations are not applicable to 
``thin'' plasmas
of only a few mean free path thickness, and more importantly they do not apply
at all to the problem of computing the angular distribution. The main
complication is that the angular distributions require a complete treatment of
all gluon and jet final state interactions. In a  direct brute force 
pQCD approach, in the {\bf thin plasma limit} where the main simplicity
arises from the fact that for a few rescatterings, 
$n_s \le 3$, the number of amplitudes is however still sufficiently small 
to allow direct computation of all interference terms.
Here we display the $n_s=1$ case, 
a complete investigation  of 
$n_s \le 3$ rescattering can be found in Ref.~\cite{GYLV99}.

The single rescattering case, $n_s=1$, was first considered by Gunion and
Bertsch~\cite{GUNION} (GB). They estimated the hadron rapidity distribution due
to gluon bremsstrahlung associated with a single elastic valence quark
scattering in the Low-Nussinov model. The most important difference compared to
the problem addressed in this paper is that GB computed the non-abelian analog
of the Bethe-Heitler formula while we are interested in the analog of radiative
energy loss in sudden processes analogous to $\beta$-decay.  The GB problem
therefore is to compute the soft radiation associated with a single scattering
of an incident on-shell quark prepared in the remote past, i.e., $t_0=-\infty$,
relative to the collision time, $t_1$.
In our case, the radiation associated with a hard jet processes in nuclear
reactions must take into account the fact that the jet parton rescatters within
a short time , $t_1-t_0 \sim R_A$, after it is produced.  The sudden appearance
of the jet color dipole moment within a time interval $\sim 1/E$ results in a
broad angular distribution of gluons even with no final state scattering.  The
induced radiation cased by rescattering necessarily interferes strongly with
this hard radiation.

To model the scattering in a thin plasma, we employ the same model of the
plasma as considered in Gyulassy--Wang (GW) \cite{MGXW}.  The scattering
centers are approximated by static color-screened potentials 
with Fourier components
$V(\vec{q})={4 \pi \alpha_s}/{(\vec{q}^2+\mu^2)}$ and
color screening mass, $\mu=4 \pi \alpha_s T^2$.

\section{Angular Distribution with One Rescattering}

The amplitude for the production of a hard jet with
momentum $P_0$ localized initially near $x_0^\mu=(t_0,\vec{0})$ we
denote by ${\cal{M}}_{J} = J(P)e^{i P x_0}$. 
The hard vertex is localized within a distance $\sim 1/P_0$,
and the amplitude, $J(P)$, is assumed to vary
slowly with $P$ on the infrared screening scale $\mu$. 
This jet radiates a gluon with momentum $\vk$, which can scatter on $n_s=1$ 
scattering center.

\begin{center}
\vspace*{9.0cm}
\includegraphics{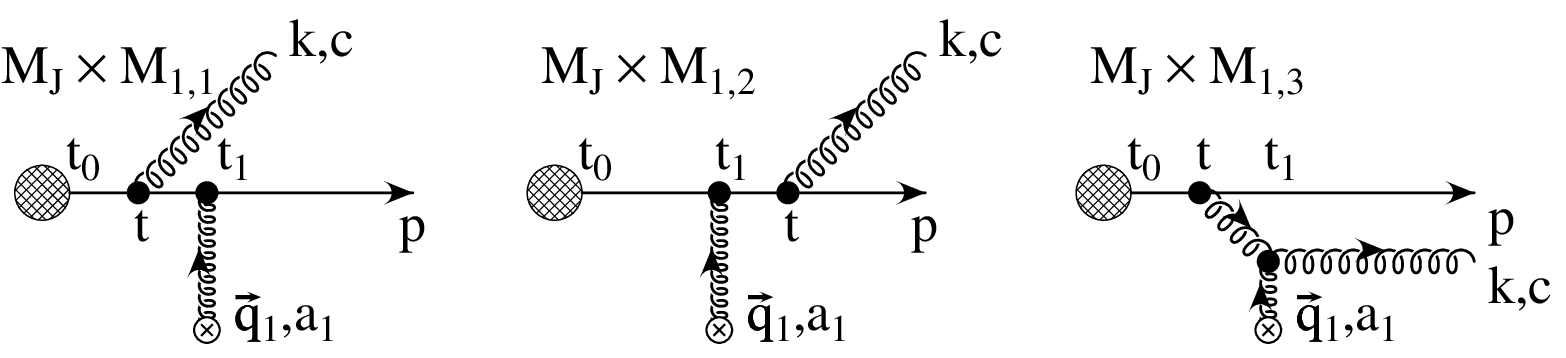}
\vskip -165pt
\begin{minipage}[t]{15.0cm}
{\small {\bf Fig.~1.}
Three contributions to the soft gluon radiation amplitude 
 {\cal M}$_J \otimes$ {\cal M}$_{1}$.}
\end{minipage}
\end{center}

The conditional double inclusive jet and gluon
probability distribution is the following:
\begin{equation}
d^{\, 6}{\cal D}^{(n_s=1)}_{rad}=\rho^{(n_s=1)}_{rad}(k,p)
\frac{d^3\vk}{(2\pi)^3 2\omega}\frac{d^3\vp}{(2\pi)^3 2p_0} =
\left \langle |{{\cal M}_J} \otimes{\cal{M}}_{n_s=1} |^2
\right \rangle 
\frac{d^3\vk}{(2\pi)^3 2\omega}\frac{d^3\vp}{(2\pi)^3 2p_0} \ \ ,
\end{equation}
which involves averaging over all initial target colors and summing over
all final colors and the locations $x_i$ of the target centers.
After averaging one obtains: 
\begin{eqnarray}
\rho^{(1)}_{rad}(k,p)&\approx &  |J(p+k)|^2 \int \frac{d^2 \vq_1}{(2\pi)^2} \,
\frac{d^2 \vq_1^{\; \prime}}{(2\pi)^2}  V(\vq_1) V^*(\vq_1^{\; \prime})
\frac{C_2(i)}{D_A}T(\vq_{1 \perp}-\vq_{1 \perp }^{\; \prime} )  
\nonumber \\[.5ex]
&\times &  {\displaystyle \left \langle \begin{array}{c}
 \tr\, \left(  \sum\limits_{i,j=1}^3 {\cal M}_{1,i}
(k,p;\vq_{1 \perp}) {\cal M}_{1,j}^\dagger (k,p;\vq_{1 \perp}^{\; \prime}  )
\right )  \end{array}  \right \rangle_t } \; .\qquad \label{rho1}
\end{eqnarray}
The main source of non-diagonal dependence arises through the
interference terms in ${\cal M}_1 {\cal M}_1^\dagger$ and the
$T(\vq_{1 \perp}-\vq_{1 \perp }^{\; \prime} )$ transverse form factor.

Following Ref.~\cite{BDMPS}, the three radiative
amplitudes {\cal M}$_{1,i}$ of Fig.1. are given by 
\begin{eqnarray}
{\cal{M}}_{1,1} &=& 2 i g_s {\vec{\epsilon}_\perp \cdot
\vk_\perp \over k^2_\perp }
(e^{i t_1 {k^2_\perp \over 2\omega}}
-e^{i t_0 {k^2_\perp \over 2\omega}} ) a_1 c \; , 
\ \ \ \ \ \ 
{\cal{M}}_{1,2} = 2 i g_s {\vec{\epsilon}_\perp \cdot
\vk_\perp \over k^2_\perp }
( -e^{i t_1 {k^2_\perp \over 2\omega}} ) c  a_1\; , \nonumber \\
{\cal{M}}_{1,3} &=& 2 i g_s {\vec{\epsilon}_\perp \cdot
(\vk - \vq_1 )_\perp \over (k - q_1)^2_\perp }
e^{i t_1 {k^2_\perp - (k - q_1)^2_\perp \over 2\omega}} \times 
  (e^{i t_1 {(k - q_1)^2_\perp \over 2\omega}}
-e^{i t_0 {(k - q_1)^2_\perp \over 2\omega}} )
 \left [ c, a_1 \right ] \; .
\label{m1}
\end{eqnarray}
The momentum distribution of
gluon radiation can be derived from the conditional
probability distribution of radiation
$\rho_{rad}^{(1)}(p,k)$, normalized by the elastic scattering
$\rho_{el}^{(1)}(p)$ \cite{GYLV99}. Before averaging
on scattering centers one can obtain from eq.~(\ref{rho1}):
\begin{eqnarray}
\frac{dN_g^{(1)}}{dy d^2\vk_\perp} =
\frac{\rho_{rad}^{(1)}(p,k)}{\rho_{el}^{(1)}(p)}&\Longrightarrow&
C_R {\alpha_s \over \pi^2}
\left\{ \vec{H}^2 + R\,(\vec{B}^2_1 +t(\vb_1)\; \vec{C}^2_1  )
- R\left ( \vec{H} \cdot \vec{B}_1 \cos (t_{10}\omega_0)
\right) \right.  \nonumber \\
&& -  R\,t(\vb_1/2 ) \left. \left(\vec{H}\cdot\vec{C}_1
\cos (t_{10} \omega_{10}) - 2\vec{C}_1\cdot\vec{B}_1
\cos (t_{10}\omega_1) \right) \right\} \,  \; . \label{dn10}
\end{eqnarray}
Here we rearranged the radiation amplitudes and 
introduced the hard term $\vH = \vk_\perp/k^2_\perp$,
the cascading term $\vC_1 = (\vk-\vq_1)_\perp/(k-q_1)^2_\perp$,
and the GB term $\vB_1=\vH-\vC_1$.
The interference terms contain phase factors composed 
from transverse energy transfer terms 
$\omega_0\equiv k^2_\perp/2\omega$, $\omega_1\equiv (k-q_1)^2_\perp/2\omega$,
$\omega_{10} \equiv \omega_1 - \omega_0$ and 
time difference term $t_{10} \equiv t_1 -t_0$. These interference terms
are responsible for ``quantum cascading'': without them one can recover
the classical cascading limit in eq.~(\ref{dn10}); with them a
complex behavior of ``quantum cascading'' dominated by these interference terms appears. 
The transverse form factor $t(\vb)=T(\vb)/T(0)$ connects to
the target size and it regulates automatically  
the gluon cascading on the finite size target object.

In the limit $t_1 \rightarrow +\infty$ the hard radiation
formula for the no-scattering case, $\vH^2$ is recovered. 
The limit $t_0 \rightarrow - \infty$ is corresponding
to the isolated bremsstrahlung amplitude and it leads to the 
characteristic  GB radiation spectrum, 
$\vH^2 + R \vB^2_1$ \cite{GUNION}.

At finite $t_1$ and $t_0=0$ the numerical evaluation of the
eq.~(\ref{rho1}) leads to the angular distribution of the radiated gluon.
Fig. 2. displays our numerical results,
normalizing the one gluon radiation  one scattering distribution,
${dN_g^{(1)}}/{dy d^2\vk_\perp}$, by the one gluon radiation 
no scattering distribution, ${dN_g^{(0)}}/{dy d^2\vk_\perp}$.
Fig.2a, displays the contribution of the time-independent, classical cascading
terms and their sum. Fig.2b, shows the strong influence of the
interference terms. Especially  at small $k_\perp$ the hard radiation 
limit is recovered through the suppression of the GB contribution.
This can be proved exactly in eq.(\ref{dn10})
by means  of $k_\perp \rightarrow 0$.
At large $k_\perp$ kinematical cuts appear from the appropriate jet energy,
however the enhancement connecting the presence of 
the $\vC_1$ cascading term can be seen clearly.

In the case of more scattering ($n_s>1$) 
the existence of $\vC_i$ cascading terms leads to 
approximate exponentiation of the relative gluon  radiation
distribution. Naive interpretation of this result would suggest that
the energy loss exponentiates with the length. 
However, in that case multiple emission processes have to be also
considered and this necessitates a renormalization of the wavefunction
as discussed in detail in Ref.~\cite{GYLV99}.
With that renormalization one finds approximately constant $dE/dx$ for
$n_s \leq 3$ up to L $\sim$ 3 fm.

\newpage
\begin{center}
\vspace*{9.0cm}
\includegraphics{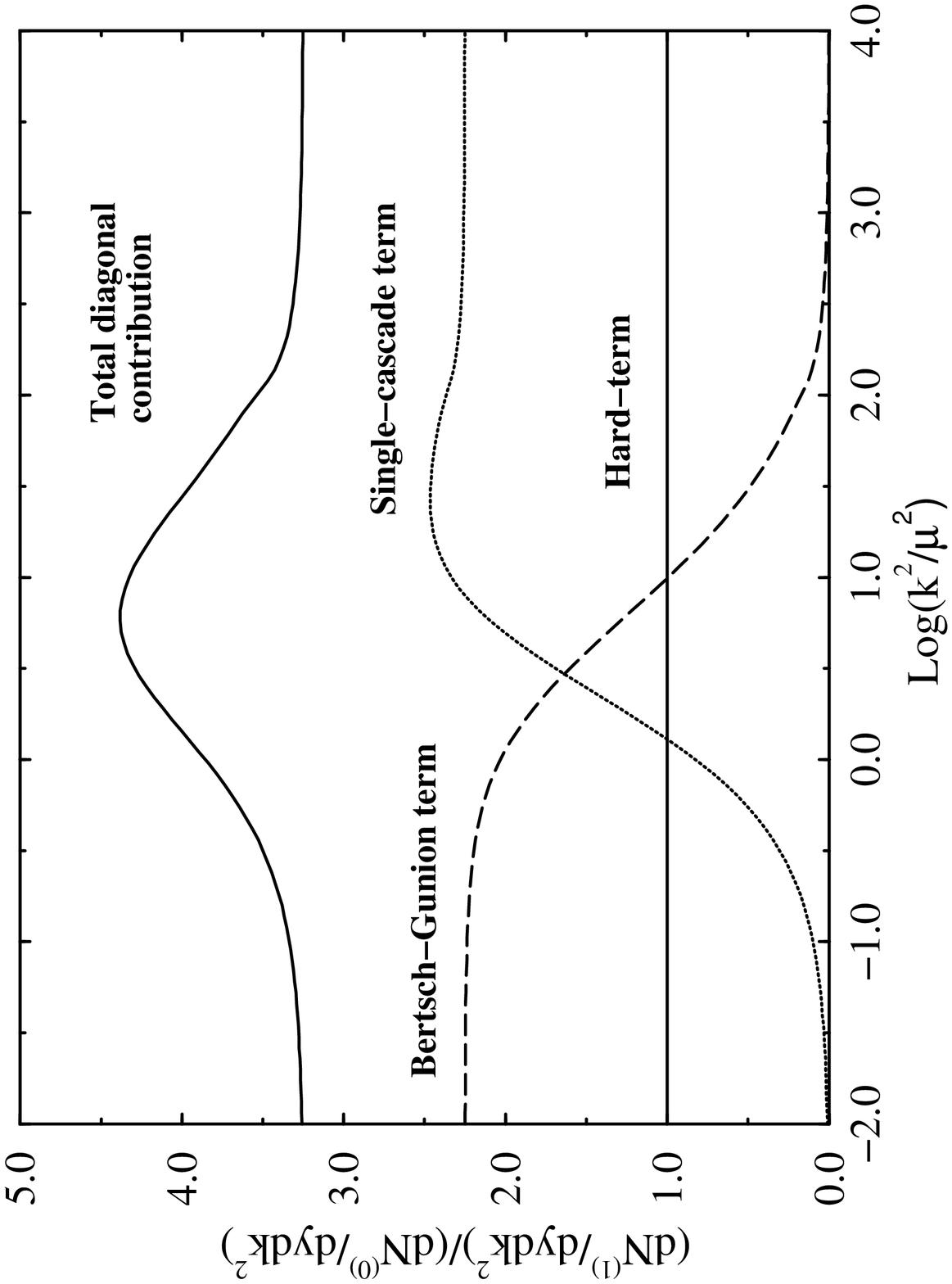}
\includegraphics{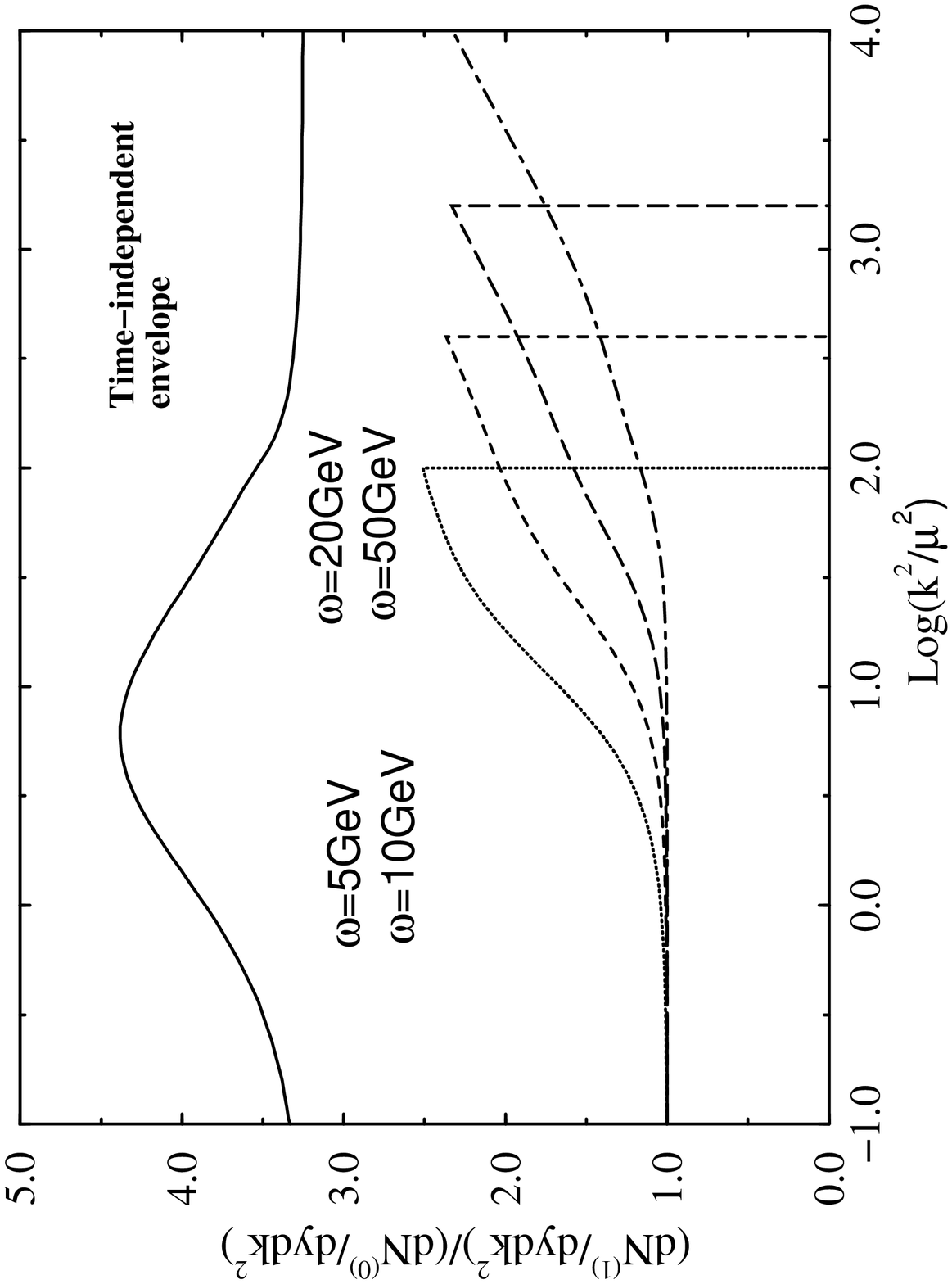}
\vskip -45pt
\begin{minipage}[t]{13.7cm}
{\small {\bf Fig.~2.}  (a) The normalized time-independent envelopes and
separate contributions to the conditional
probability distribution of gluons associated with
a single rescattering $n_s=1$ of quark jet (${\mbox E}_{\mbox
{CM}}=100{\mbox {GeV}}$) is shown
as a function of the logarithmic ``angle'',
$\log_{10} k_\perp^2/\mu^2$ with $\mu=0.5$ GeV.
These curves correspond to the incoherent parton cascade limit.
(b) For finite $t_{10}=1$ fm/c case destructive interference
limits the corrections to the hard self-quench
distribution to higher angles as shown for different gluon energies.
}
\end{minipage}
\end{center}

{\bf Acknowledgment:}
We thank Urs Wiedemann for many discussions.
This work was supported by the DOE Research Grant under Contract
No. De-FG-02-92ER-40764, by the US-Hungarian Joint Fund No.652
and OTKA No.T025579, T024094.

\end{document}